\documentclass[11pt]{article}
\usepackage{moriond,epsfig}

\def\be{\begin{equation}}
\def\ee{\end{equation}}
\def\bea{\begin{eqnarray}}
\def\eea{\end{eqnarray}}

\begin{document}
\vspace*{4cm}
\title{MEASUREMENT OF THE PION FORMFACTOR WITH KLOE
\\
AND STUDY OF THE REACTION
$f_0(980)\to\pi^+\pi^-$}

\author{The KLOE collaboration
~\footnote{
The KLOE Collaboration: F.~Ambrosino,
A.~Antonelli,
M.~Antonelli,
C.~Bacci,
P.~Beltrame,
G.~Bencivenni,
S.~Bertolucci,
C.~Bini,
C.~Bloise,
S.~Bocchetta,
V.~Bocci,
F.~Bossi,
D.~Bowring,
P.~Branchini,
R.~Caloi,
P.~Campana,
G.~Capon,
T.~Capussela,
F.~Ceradini,
S.~Chi,
G.~Chiefari,
P.~Ciambrone,
S.~Conetti,
E.~De~Lucia,
A.~De~Santis,
P.~De~Simone,
G.~De~Zorzi,
S.~Dell'Agnello,
A.~Denig,
A.~Di~Domenico,
C.~Di~Donato,
S.~Di~Falco,
B.~Di~Micco,
A.~Doria,
M.~Dreucci,
G.~Felici,
A.~Ferrari,
M.~L.~Ferrer,
G.~Finocchiaro,
S.~Fiore,
C.~Forti,
P.~Franzini,
C.~Gatti,
P.~Gauzzi,
S.~Giovannella,
E.~Gorini,
E.~Graziani,
M.~Incagli,
W.~Kluge,
V.~Kulikov,
F.~Lacava,
G.~Lanfranchi,
J.~Lee-Franzini,
D.~Leone,
M.~Martini,
P.~Massarotti,
W.~Mei,
L.~Meola,
S.~Miscetti,
M.~Moulson,
S.~M\"uller,
F.~Murtas,
M.~Napolitano,
F.~Nguyen,
M.~Palutan,
E.~Pasqualucci,
A.~Passeri,
V.~Patera,
F.~Perfetto,
L.~Pontecorvo,
M.~Primavera,
P.~Santangelo,
E.~Santovetti,
G.~Saracino,
B.~Sciascia,
A.~Sciubba,
F.~Scuri,
I.~Sfiligoi,
T.~Spadaro,
M.~Testa,
L.~Tortora,
P.~Valente,
B.~Valeriani,
G.~Venanzoni,
S.~Veneziano,
A.~Ventura,
R.~Versaci,
G.~Xu.}
\\represented by Achim Denig}
\address{Universit\"at Karlsruhe, IEKP, Postfach 3640, 76021 Karlsruhe, Germany}

\maketitle\abstracts{
At the Frascati $\phi$-factory DA$\Phi$NE the pion formfactor is measured
by means of the 'radiative return', i.e. by using events in which one
of the collider electrons (positrons) has radiated an initial state radiation 
photon, lowering in such a way the invariant mass ${\rm M}_{\pi\pi}$ of the two-pion-system. 
In a recent publication of the KLOE collaboration the initial state radiation
photon had been required to be at small polar angles with respect to the
beam axis. We are presenting results from a new and complementary analysis in which the 
photon is tagged at large polar angles. Only like this the threshold region
${\rm M}_{\pi\pi}^2<0.35 {\rm GeV}^2$ becomes accessible. Moreover, the final state $\pi^+\pi^-\gamma$
allows to study the $\phi$ radiative decay into
the scalar particle $f_0(980)$ with $f_0(980)\to\pi^+\pi^-$. For the first
time the two-pion mass spectrum could be fitted with different 
theoretical models for the description of this $\phi$ radiative decay.
}

\section{The radiative return and its connection to the muon anomaly}
Precision measurements of the cross section for $e^+e^-$ annihilation into hadrons
are of utmost importance for an interpretation of the recent measurement of
the anomalous magnetic moment of the muon $a_\mu$ 
at the Brookhaven National Laboratory (experiment E821, $0.5$ ppm accuracy)~\cite{brookh}. 
A comparison of the E821 measurement with the theory prediction 
allows a unique test of the standard model of particle physics.
The theory prediction is however
limited by the hadronic contribution $a_\mu^{\rm hadr}$ and can only be derived by means of a 
dispersion integral, using hadronic cross section data as input~\cite{ej}~\cite{dh}.
The process $e^+e^-\to\pi^+\pi^-$ below 1 GeV is of special 
importance since it contributes to $\sim 60\%$ to the total integral.\\
Recently it could be shown that at particle factories, such as DA$\Phi$NE or 
the B-factories, which are operated at a constant center-of-mass energy $\sqrt{s}$,
the hadronic cross section becomes accessible over a wide energy range $<\sqrt{s}$ using
events with initial state radiation (ISR), lowering in such a way the
invariant mass of the hadronic system ${\rm M}_{\rm hadr}$ (so-called 'radiative return'). 
The KLOE collaboration could prove for the two-pion-channel (pion formfactor) 
that this method is not only complementary but also
competitive with the standard energy scan method~\cite{kloe}. The results of the
published KLOE measurement, a comparison with recent precision data from the VEPP-2M
collider in Novosibirsk and the implications for
the muon anomaly can be summarized as follows:
\begin{itemize}
\item{In the published KLOE analysis events have been selected, in which the ISR-photon is 
emitted at small (large) polar angles $\Theta_\gamma <15^{\rm o}$ and $\Theta_\gamma>165^{\rm o}$
with respect to the beam axis. 
The photon cannot be tagged in such an approach (so-called {\it untagged analysis}).}
\item{From the radiative cross section $e^+e^-\to\pi^+\pi^-\gamma$ the non-radiative
cross section $e^+e^-\to\pi^+\pi^-$ (pion formfactor) has been extracted
using a radiator function (obtained from the Monte-Carlo generator PHOKHARA~\cite{phok}) 
for the theoretical description of the ISR-process. The total 
systematic error of $1.3\%$ for the pion formfactor includes contributions from
the experimental side (e.g. efficiencies, background, in total $0.9\%$) and from the theory
side (e.g. radiator function, Bhabha cross section for normalization, in total $0.9\%$).}
\item{The pion formfactor measured by KLOE is in fair agreement with measurements
coming from the Novosibirsk collider VEPP-2M (experiments CMD-2~\cite{cmd2} and SND~\cite{snd}), 
where an energy scan has been performed.
Please notice that the SND data have come closer 
to the KLOE spectrum at high masses above the $\rho$ peak,
only after the correct treatment of the radiative corrections in the SND analysis, 
even though a systematic shift of few percent is still visible. }
\item{All three experiments CMD-2, KLOE and SND show large
deviations of up to $15\%$ in the mass range above the $\rho$ peak with respect to spectral functions 
obtained from hadronic $\tau$ decays,
which can be related to the $\pi^+\pi^-$ cross section by means of the conserved vector current
(CVC) theorem and after correcting for isospin breaking effects. 
The origin of the deviation between $e^+e^-$- and $\tau$-data is not understood.}
\item{The contribution of the two-pion channel to the dispersion integral for $a_\mu^{\rm hadr}$
has been computed with KLOE data. In the mass range
$0.35 < {{\rm M}_{\pi\pi}^2} < 0.95 {\rm GeV}^2$ the value for $a_\mu^{\pi\pi}$ is 
$(388.7 \pm 0.8_{\rm stat} \pm 3.5_{\rm syst} \pm 3.5_{\rm theo}){\rm x}10^{-10}$. 
CMD-2 and SND agree in the dispersion integral within $0.5$ standard deviations 
with the KLOE measurement~\footnote{in the somewhat smaller range $0.37 < {{\rm M}_{\pi\pi}^2} 
< 0.93 {\rm GeV}^2$}.}
\end{itemize}

\section{Radiative return with tagged photons} 
The analysis described above, in which the ISR-photon is emitted at small polar angles, does
not allow to cover the threshold region ${\rm M}_{\pi\pi}^2 < 0.35 {\rm GeV}^2$, since in this kinematical
region the two tracks are emitted essentially back-to-back to the ISR-photon and hence cannot be
detected simultaneously in the fiducial volume 
defined for the pion tracks $50^{\rm o}<\Theta_\pi<130^{\rm o}$. In order to
measure the pion formfactor at threshold, we are now performing a complementary
analysis, in which the ISR-photon is tagged at large polar angles $50^{\rm o}<\Theta_\gamma<130^{\rm o}$.
Due to the $1/s^2$ dependence in the dispersion integral for $a_\mu^{\rm hadr}$,
the low mass region of the two-pion cross section is actually giving a $\sim 20\%$ contribution
to the total integral and hence an improved determination of the cross section at threshold is 
needed. 
The tagged photon
analysis allows also a very valuable cross check of the $\rho$ peak region, which was covered in the
published small angle analysis. It should be noted that
new analysis tools are used and that the radiator function and FSR corrections are 
different. In the following we comment on the status of the analysis and the specific
requirements for the tagged analysis.
\begin{figure}[ht]
  \centering
  \epsfig{file=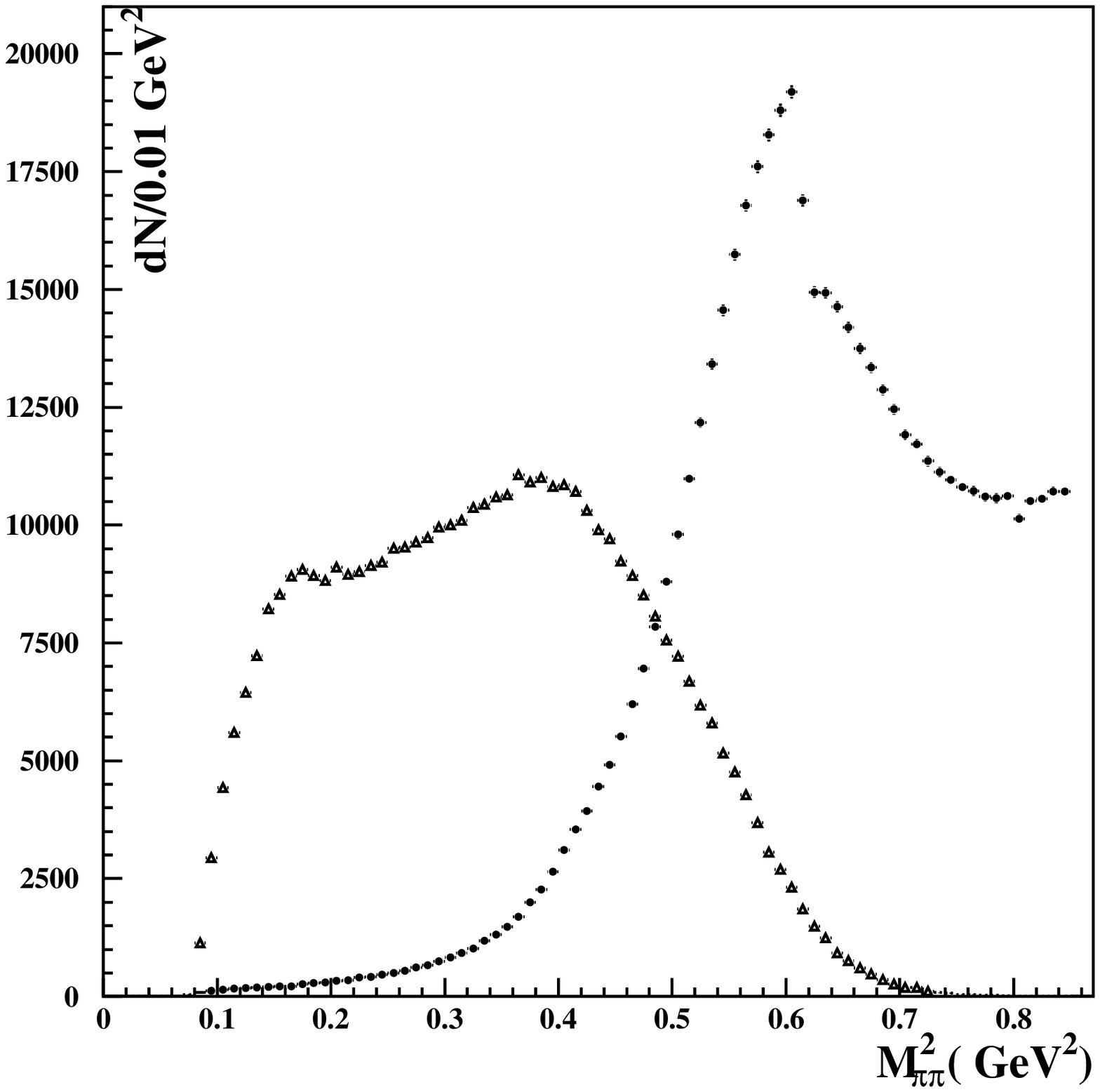, width=6cm}
\hspace{1.3cm}
  \epsfig{file=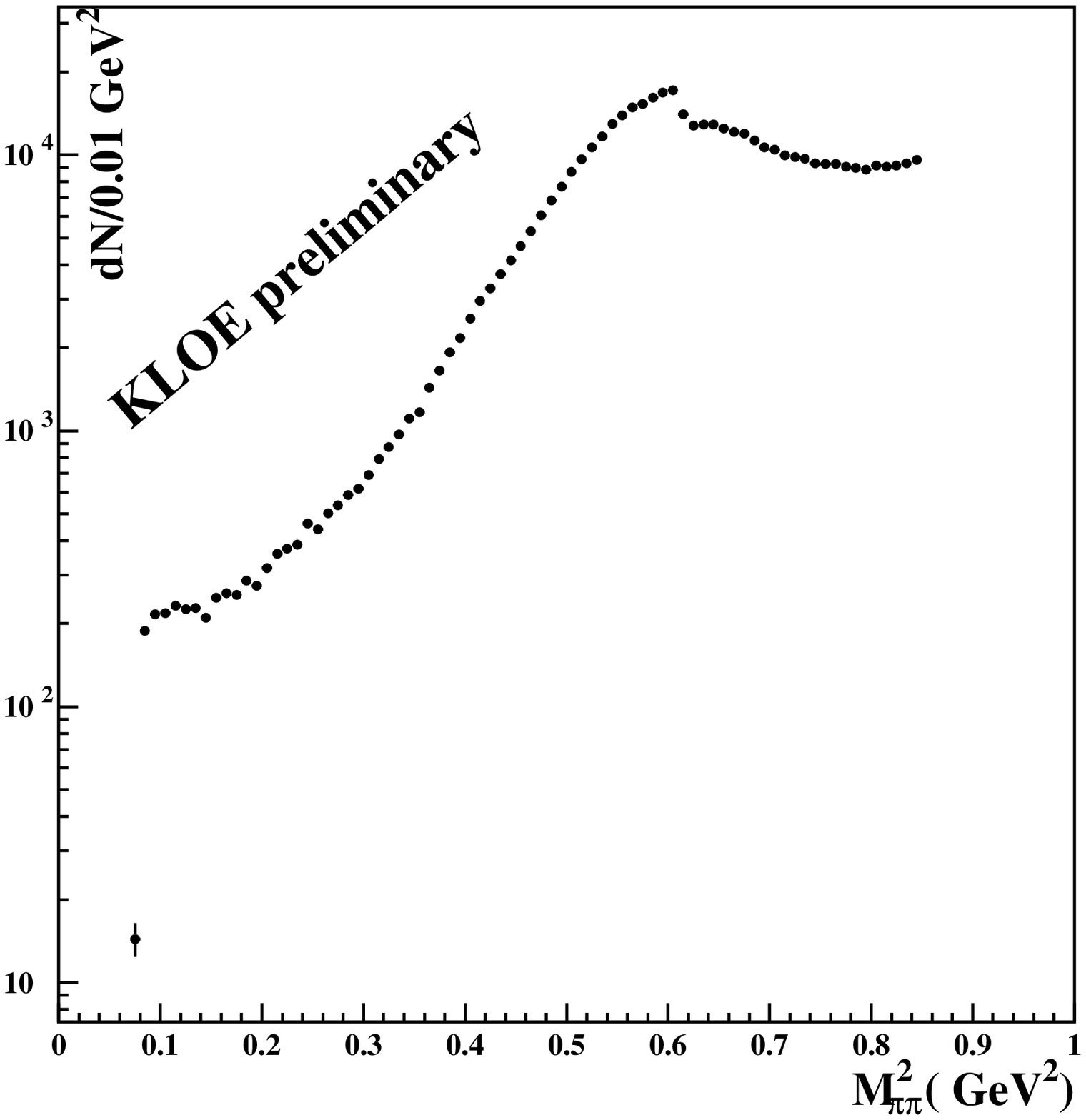, width=5.6cm}
  \caption{\small{\it Left: Mass spectrum for signal events $\pi^+\pi^-\gamma$ (circles) 
as expected by Monte-Carlo simulation after the
application of the acceptance cuts. Background from $\phi \to \pi^+\pi^-\pi^0$ 
(triangles) is dominating the low mass region. Right: Mass spectrum from data after the full analysis chain.
The large background from $\phi \to \pi^+\pi^-\pi^0$ can be removed by means of dedicated selection cuts.}}
  \label{data}
\end{figure}
\begin{itemize}
\item{The cross section for large angle photon events is approximately a factor $5$ smaller
than for small angle events. Especially in the low ${\rm M}_{\pi\pi}$ region, where the pion
formfactor is getting rather small, statistics becomes an issue. We use the data taken 
in 2002, which corresponds to an integrated luminosity of $\sim 240 {\rm pb}^{-1}$. Like this
we select $\sim 45 000$ large angle photon events.}
\item{At large photon angles background from $\phi\to\pi^+\pi^-\pi^0$ is large, 
especially in the low ${\rm M}_{\pi\pi}$ region as can be seen in fig.~\ref{data}(left). However, due 
to the fact that the ISR-photon is tagged and the event kinematics is closed, it is 
possible to apply stringent cuts to 
reduce this kind of background. We ask the angle of the missing momentum - calculated 
from the two pion tracks - to be within a certain window close to the angle of the
tagged photon. For $\pi^+\pi^-\gamma$ events this angle is expected to be $\sim 0^{\rm o}$, while for
$\pi^+\pi^-\pi^0$ events it has a broad distribution peaked at $\Omega \approx 20^{\rm o}$.
Moreover, we perform a kinematic fit in the hypothesis of the background
channel $\pi^+\pi^-\pi^0$, using four-momentum conservation and the $\pi^0$ mass as
constraints, and cut on $\chi^2_{\pi\pi\pi}$.}
\item{Another important background arises from events, in which the photon is not emitted
from the initial electrons or positrons, but from the pions (final state radiation, FSR). 
In the radiative return these
kind of events are a background and have to be cut from Monte-Carlo simulation. We use the
PHOKHARA code~\cite{phok}, in which the model of scalar QED (pointlike pions) 
is used for the estimate of FSR, and in which FSR-corrections up to NLO are considered. 
Within this model we find a contribution from FSR at large 
${\rm M}_{\pi\pi}$ of up to $20\%$. The model assumption of scalar 
QED can be tested by means of the 
forward-backward asymmetry (see below), which arises from the ISR-FSR-interference and has been found to 
describe well the data with a precision of better than few percent.}
\item{Beneath FSR, a further irreducible background is coming from the $\phi$ radiative 
decay $\phi\to f_0(980)\gamma \to \pi^+\pi^-\gamma$. Also in this case we deduce the 
contribution from Monte-Carlo simulation. 
This $\phi$ decay is very interesting in itself and 
we report in the following section on a recent KLOE fit~\cite{cesare} of 
the mass spectrum for $f_0(980) \to \pi^+\pi^-$, using
different theoretical models for the decay amplitude as well as a measurement of the
charge asymmetry.}
\end{itemize}  
The analysis using tagged photons is in an advanced state. 
As in the published small photon angle analysis, we are studying 
the selection efficiencies (trigger, tracking, vertex, photon detection) directly from
data using independent control samples. We find an overall good agreement
with the predictions from simulation. While the trigger inefficiency is below $0.1\%$, 
we find values for the vertex efficiency and tracking efficiency of $99.2\%$ and
$\sim 98\%$, respectively. The total selection efficiency (not taking into account the
geometrical acceptance) is $80-90\%$ and rather flat in ${\rm M}_{\pi\pi}$.
\\
The $\pi^+\pi^-\gamma$ event yield after application of all selection cuts is shown in
fig.~\ref{data}(right). As can be seen, the threshold region is covered in the large angle analysis 
and the huge background
from $\phi\to\pi^+\pi^-\pi^0$ can be considerably reduced. No corrections for
$\phi\to f_0(980)\gamma \to \pi^+\pi^-\gamma$ are yet applied in this plot, while
FSR corrections are taken into account and the background from $\mu^+\mu^-\gamma$ and $\pi^+\pi^-\pi^0$
is subtracted.
\\
As stressed above, the main limitation of the large angle analysis is due to the
background associated with the $\phi$-decays into $\pi^+\pi^-\pi^0$ and to
$f_0(980)\gamma$. In order
to reduce the systematic errors associated to these channels to a very low level, 
the DA$\Phi$NE collider has recently taken 
data off-resonance at a center-of-mass energy of $\sqrt{s}=1.00 {\rm GeV}$
(December 2005 to March 2006, $250 {\rm pb}^{-1}$ integrated luminosity). 
The analysis of these data will allow an improved determination of
the threshold region. Moreover, together with the data taken on-peak it will be possible to study 
the interference of the $f_0(980)$ amplitude with FSR.
\begin{figure}[ht]
  \centering
  \epsfig{file=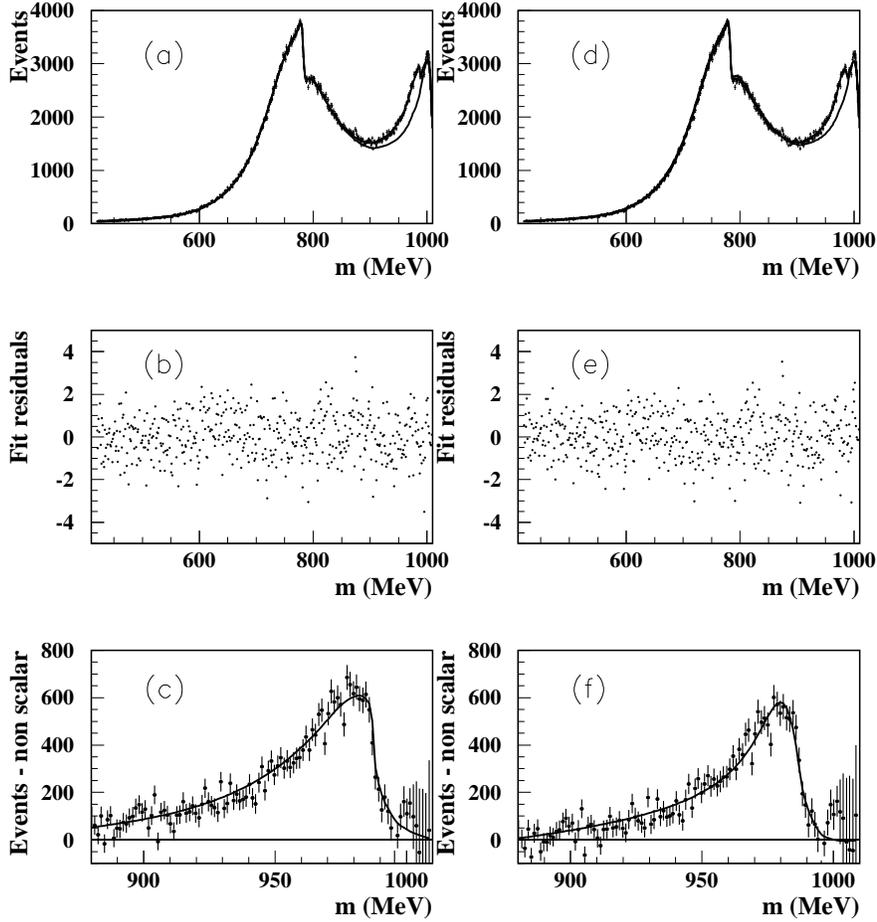,width=12.5cm}
  \caption{\small{\it Result of the KL fit (a)-(b)-(c) and of 
      the NS fit (d)-(e)-(f). (a)-(d) Data spectrum
      compared with the fitting function (upper curve following the data points)
      and with the estimated non-scalar part
      of the function (lower curve); (b)-(e) fit residuals as a function of $m$; 
      (c)-(f) the fitting function is compared to the spectrum
      obtained subtracting to the measured data the non-scalar 
      part of the function in the f$_0$ region. 
  }}
  \label{fitresults}
\end{figure}

\section{Study of the reaction $f_0(980)\to\pi^+\pi^-$}
The scalar mesons $f_0(980)$ and $a_0(980)$ are produced at DA$\Phi$NE in radiative
decays of the $\phi$ meson and the study of these decays has caused much attention
due to the potential sensitivity to distinguish between different models for the nature
of the scalars ($q{\bar q}$, KK molecule, 4-quark)~\cite{f0}. In the charged final state  
$\phi \to f_0(980)\gamma \to \pi^+\pi^-\gamma$ the main background is arising from
ISR and FSR continuum events $e^+e^-\to\pi^+\pi^-\gamma$. 
In order to reduce as much as possible the ISR contribution
and in order to enhance the relative amount of the $f_0(980)$ signal, the photon is 
required to be detected at large polar angles. The selection is therefore identical 
to the large photon angle analysis discussed above.
\\ 
The ${\rm M}_{\pi\pi}$ spectrum
is fitted with a function assuming the following contributions: $f_0(980)$, ISR, FSR, 
FSR-$f_0(980)$-interference and $\rho\pi$, where the latter two arise from the
interference between the FSR- and the $f_0(980)$-amplitude and from the decay chain
$\phi \to \rho\pi \to \pi^+\pi^-\gamma$, respectively. 
The fit clearly prefers a negative interference between FSR and the scalar amplitude, for
which three theoretical models have been used: (i) the kaon-loop (KL) model 
described in ref.~\cite{kl}, (ii) the no-structure (NS) model of ref.~\cite{ns} and (iii) the scattering-amplitude
model (SA) of ref.~\cite{sa}. Further details concerning these models can be found in the appropriate
references. Free parameters concerning the scalar amplitudes are the mass of the $f_0(980)$,
the couplings $f_0 K^+K^-$, $f_0 \pi^+\pi^-$ (for KL and NS) and the direct coupling
$\phi f_0 \gamma$ (for NS). In the case of the SA model, in which the scalar amplitude is the
sum of the $\pi\pi\to\pi\pi$ and $\pi\pi\to K K$ scattering amplitudes, the
fit parameters are an overall phase and the 6 coefficients of an expansion, which 
parametrize the mass dependence of the two scattering amplitudes.
\\
We fit the data in the region $420<{\rm M}_{\pi\pi}<1010$ MeV using bins 
$1.2$ MeV wide. The results of the fit are shown in fig.~\ref{fitresults} for the KL and NS models.
The $f_0(980)$ signal appears as an excess of events in the region between 900 and 1000 MeV.
An attempt to include a second scalar meson, namely the $f_0(600)$, does not improve the fit quality. 
After subtraction of the non-scalar part, an asymmetric peak around $980$ MeV with
a FWHM of $30-50$ MeV is obtained, as shown in fig.~\ref{fitresults} in the lower row. Such a peak
does not directly represent the $f_0(980)$ shape but it results from the sum of the broad
signal amplitude and the negative interference term that cancels the low mass
tail. The ratio of the $f_0 K^+K^-$ and $f_0 \pi^+\pi^-$ couplings 
$R=g^2_{f_0  K^+K^-}/g^2_{f_0  \pi^+\pi^-}$ is well above 2 (KL: $2.2-2.8$, NS: $2.6-4.4$), 
indicating a strong coupling of the $f_0(980)$ to strangeness.
For the SA fit, which has a worse fit quality, we find only marginal agreement with 
the KL and NS results. Especially the branching ratio 
${\rm BR}(\phi \to f_0 \gamma){\rm x}{\rm BR}(f_0 \to \pi^+\pi^-)$ can be extracted and is found to be
in the order $10^{-5}$, being about one order of magnitude lower than the values
extracted from the KL and NS parameters. The complete list of fit results and further 
information concerning the procedure can be found in ref.~\cite{cesare}.
\begin{figure}[ht]
  \centering
  \epsfig{file=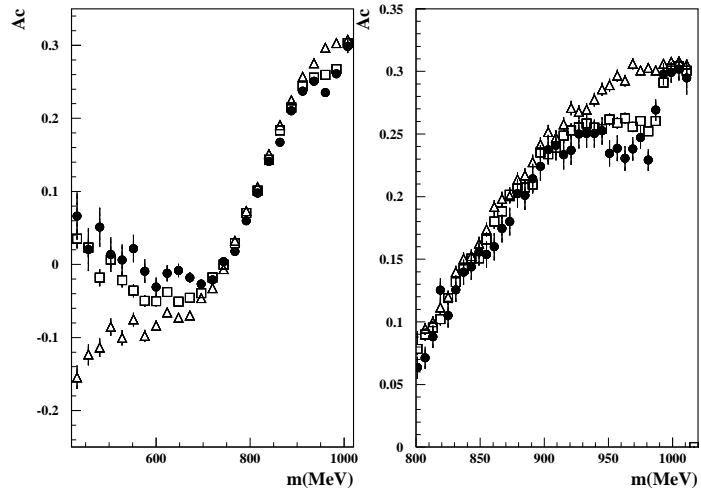,width=10cm}
  \caption{\small{\it The forward-backward asymmetry for data (full circles) 
      compared to the Monte-Carlo expectations based on the non-scalar part of
      the spectrum only (open triangles), and on the non-scalar plus $f_0(980)$ part
      obtained from the KL amplitude(open squares). 
      The right plot shows the detail of the comparison in the
      f$_0$ mass region. 
  }}
  \label{comparison}
\end{figure}
\\
As proposed in ref.~\cite{czyz}, the behaviour of the forward-backward asymmetry as a function of 
${\rm M}_{\pi\pi}$ has been studied. The forward-backward asymmetry is defined as
\begin{equation}
A_{c}={{N(\theta_{\pi^+}>90^{\circ})-N(\theta_{\pi^+}<90^{\circ})}\over
 {N(\theta_{\pi^+}>90^{\circ})+N(\theta_{\pi^+}<90^{\circ})}},
\label{asymmetry}
\end{equation}
and is plotted in fig.~\ref{comparison} for data and for two Monte-Carlo predictions, excluding and including
the presence of the $f_0(980)$ amplitude. As can be seen from the plot, the inclusion 
of the scalar amplitude is necessary for an acceptable agreement at high and low masses. 
While at low masses the $f_0(980)$ amplitude is cancelled in the mass spectrum due to 
the destructive intereference with FSR, on the contrary the scalar amplitude is very
evident in the charge asymmetry due to the interference with ISR. 


\end{document}